\documentclass[showpacs,aps,twocolumn,superscriptaddress,preprintnumbers,nofootinbib]{revtex4-1}
\usepackage{amsmath}
\usepackage{epsfig}
\usepackage{subfigure}
\usepackage{float}
\usepackage{verbatim}
\usepackage{axodraw4j}
\usepackage{pstricks}
\usepackage{color}
\usepackage{slashed}
\usepackage{multirow}
\usepackage{pstricks}
\usepackage{color}
\usepackage{booktabs}
\usepackage{subfigure}
\usepackage{epstopdf}

\newcommand{\ie}{{\it i.e.}}

\begin{document}

%\leftline{}

\preprint{\font\fortssbx=cmssbx10 scaled \magstep2
\hbox to \hsize{
\hfill$\vcenter{\hbox{ CP3-14-84}
\hbox{ MITP/14-085}
\hbox{ IPPP/14/109}
\hbox{ DCPT/14/218}
%                \hbox{\bf hep-ph/yymmnnn}
%                \hbox{November 2013}
                }$}
}

\title{Automatic computations at next-to-leading order in QCD \\[2pt] for top-quark flavor-changing neutral processes}

\author{Celine Degrande}
\affiliation{
Institute for Particle Physics Phenomenology, Department of Physics\\
Durham University, Durham DH1 3LE, United Kingdom
}
\author{Fabio Maltoni}
\affiliation{
Centre for Cosmology, Particle Physics and Phenomenology,
Universit\'e Catholique de Louvain, B-1348 Louvain-la-Neuve, Belgium
}
\author{Jian Wang}
\affiliation{
  PRISMA Cluster of Excellence \& Mainz Institute for Theoretical Physics,
Johannes Gutenberg University, D-55099 Mainz, Germany}
\author{Cen Zhang}
\affiliation{
Department of Physics, Brookhaven National Laboratory, Upton, New York 11973, USA
}

\begin{abstract}
Computations at next-to-leading order in the Standard Model offer new technical
challenges in the presence of higher dimensional operators. We introduce a
framework that, starting from the top-quark effective field theory at dimension
six, allows one to make predictions for cross sections as well as distributions
in a fully automatic way. As an application, we present the first complete
results at next-to-leading order in QCD for flavor-changing neutral
interactions including parton shower effects, for $tZ$, $th$, $t\gamma$
associated production at the LHC.
 \end{abstract}

\pacs{14.65.Ha,12.38.Bx}

\maketitle

\section{Introduction} 
The millions of top quarks already produced at the LHC together with the tens
of millions expected in the coming years will provide a unique opportunity to
search for interactions beyond the Standard Model (SM).  Among them
flavor-changing neutral (FCN) interactions are of special interest.  In the SM,
FCN interactions can be generated at one loop, yet they turn out to be
suppressed by the Glashow-Iliopoulos-Maiani mechanism~\cite{Glashow:1970gm}.
The resulting FCN decay modes of the top quark have branching ratios of order
$10^{-12}$--$10^{-15}$ \cite{Mele:1998ag,Eilam:1990zc,AguilarSaavedra:2002ns}.
Thus, any signal for top-quark FCN processes at a measurable rate would
immediately indicate new physics in the top-quark sector.  These processes have
been searched for already at different colliders, including LEP2, HERA,
Tevatron and more recently at the LHC~\cite{Agashe:2014kda}.  So far no signal
has been observed and limits have been set on the coupling strengths.  

The most important top-quark FCN processes at the LHC include decay processes
such as $t\to qB$ and production processes such as $qg\to t$ and $qg\to tB$,
where $q$  is a $u$ or $c$ quark and $B$ is a neutral boson, \ie,
$B=\gamma,Z,h$.  In general, the decay processes are equally sensitive to $utB$
and $ctB$ couplings, while the production modes are less sensitive to $ctB$,
but may provide a better handle on a certain class of $utB$ couplings
\cite{AguilarSaavedra:2004wm}.  In addition, compared to decay modes,
single-top production can provide more information. First, it makes a wider
range of  kinematic variables accessible, helping in the discrimination of the
light quark flavors involved and the structure of the $qtB$ couplings.  Second,
it probes interactions at higher scales where new physics effects could be
enhanced. In general, being somewhat complementary, both decay and production
processes are used for setting the most stringent constraints.

Leading order (LO) predictions for the production processes suffer from large
uncertainties due to missing higher order corrections.  To curb such
uncertainties, next-to-leading order (NLO) predictions in QCD for this class of
processes have started to be calculated in recent
years~\cite{Liu:2005dp,Gao:2009rf,Zhang:2011gh,Li:2011ek,Wang:2012gp}, providing
a much better, yet incomplete, picture of their relevance.  In general,
corrections are found to be large, of order $30\%$ to $80\%$ and to lead to
considerable reductions of  the residual theoretical uncertainties.  Both
aspects are important in bounding and possibly extracting top-quark FCN
couplings at the LHC. 

In this paper we present the first automatic computations for top-quark FCN
production processes, $qg\to tB$ with $B=\gamma,Z,h$, at NLO in QCD, by
implementing all flavor-changing dimension-six fully gauge-invariant operators
in {\sc FeynRules}~\cite{Alloul:2013bka} and then passing this information into
the {\sc MadGraph5\_aMC@NLO} framework~\cite{Alwall:2014hca}.  Compared to
previous works~\cite{Drobnak:2010wh,Drobnak:2010by,
Zhang:2011gh,Zhang:2013xya,Zhang:2014rja,Li:2011ek,Wang:2012gp},
the salient features of our results can be summarized as follows. Our study is
the first where all relevant dimension-six operators for this class of
processes (associated production with a boson) are consistently taken into
account.  In particular the vector-current like $tqZ$ coupling in $ug\to tZ$,
and the $tug$ and $tugh$ couplings in $ug\to th$ are included here for the
first time.  Second, our results are obtained via a fully automatic chain of
tools that allows one to go directly from the Lagrangian to the hard events by
performing its renormalization at one loop, and then passing the corresponding
Feynman rules to the {\sc MadGraph5\_aMC@NLO} to generate all the elements
necessary for a computation at NLO in QCD. In particular, other processes
triggered by the same set of operators can be seamlessly computed within the
same framework. Third, event generation is also automatically available at NLO
accuracy, by matching it to the parton shower via the MC@NLO
formalism~\cite{Frixione:2002ik} so that results can be employed in realistic
experimental simulations.  Finally, another important aspect of this work is
that it provides a proof of principle that fully automatic computation of cross
sections at NLO in QCD is possible in the context of the full dimension-six
Lagrangian of the SM. Higher order computations in effective field theories,
which are renormalizable only order by order in $1/\Lambda$, $\Lambda$ being
the scale of new physics, present novel technical challenges. In general, UV
divergences generated by one operator at a certain order of $1/\Lambda$ have to
be absorbed also by other effective operators.  As a result, the full set of
relevant operators together with their operator mixing effects need to be
considered simultaneously, and appropriate UV counterterms have to be
implemented in the calculation.  Our method and its implementation are fully
general and can cover arbitrary NLO calculations in the complete dimension-six
Lagrangian of the SM.

\section{Framework}
The FCN couplings of the top quark can be parametrized using either fully
gauge-symmetric dimension-six operators~\cite{AguilarSaavedra:2008zc,
AguilarSaavedra:2009mx} or dimension-four and dimension-five operators in the
electroweak broken phase~\cite{Beneke:2000hk,AguilarSaavedra:2004wm}.  The
latter approach has some intrinsic limitations~\cite{Durieux:2014xla}, and we will
use the dimension-six operators throughout the paper.  The effective Lagrangian
can be written as
\begin{equation}
  \mathcal{L}_\mathrm{EFT}=\mathcal{L}_\mathrm{SM}+\sum_i\frac{C_i}{\Lambda^2}O_i+H.c.
\end{equation}
In this work we consider $qtB$ couplings at the dimension-six level. The
relevant operators must involve one top quark and one light quark.  They are
\newcommand{\FDF}{\left(\varphi^\dagger\overleftrightarrow{D}_\mu\varphi\right)}
\newcommand{\FDFI}{\left(\varphi^\dagger\overleftrightarrow{D}^I_\mu\varphi\right)}
\begin{flalign}
    &O_{\varphi q}^{(3,i+3)}=i\FDFI(\bar{q}_i\gamma^\mu\tau^IQ)
    \nonumber\\
    &O_{\varphi q}^{(1,i+3)}=i\FDF(\bar{q}_i\gamma^\mu Q)
    \nonumber\\
    &O_{\varphi u}^{(i+3)}=i\FDF(\bar{u}_i\gamma^\mu t)
    \nonumber\\
    &O_{uB}^{(i3)}=g_Y(\bar{q}_i\sigma^{\mu\nu}t)\tilde{\varphi}B_{\mu\nu},
    \quad
    O_{uW}^{(i3)}=g_W(\bar{q}_i\sigma^{\mu\nu}\tau^It)\tilde{\varphi}W^I_{\mu\nu}
    \nonumber\\
    &O_{uG}^{(i3)}=g_s(\bar{q}_i\sigma^{\mu\nu}T^At)\tilde{\varphi}G^A_{\mu\nu},
    \quad
    O_{u\varphi}^{(i3)}=(\varphi^\dagger\varphi)(\bar{q}_it)\tilde\varphi \,,
    \nonumber
\end{flalign}
where the operator notation is consistent with
Ref.~\cite{Grzadkowski:2010es}, with additional flavor indices.  On the right
hand side, the subscript $i=1,2$ represents the generation of the light quark
fields.  $u_i$ and $q_i$ are single and doublet quark fields of the first two
generations, respectively, while $t$ and $Q$ are of the third generation.  $\varphi$
is the Higgs doublet. A diagonal CKM matrix is assumed. The group generators
are normalized such that
$\textrm{Tr}\left(T^AT^B\right)=\delta^{AB}/2$ and
$\textrm{Tr}\left(\tau^I\tau^J\right)=2\delta^{IJ}$, and
$\varphi^\dagger\overleftrightarrow{D}_\mu\varphi \equiv
\varphi^\dagger{D}_\mu\varphi-{D}_\mu\varphi^\dagger\varphi$,
$\varphi^\dagger\overleftrightarrow{D}_\mu^I\varphi \equiv
\varphi^\dagger\tau^I{D}_\mu\varphi-{D}_\mu\varphi^\dagger\tau^I\varphi$.
For operators with $(i3)$ superscript, a similar set of operators with $(3i)$
flavor structure can be obtained by interchanging $(i3)\leftrightarrow (3i),\
t\leftrightarrow u_i$ and $Q\leftrightarrow q_i$.  The first three operators
give rise to $V/A$ couplings of $Z$ to a flavor-changing current, which were not
considered in previous calculations of Ref.~\cite{Li:2011ek}.  The
$O_{uB}^{(i3,3i)}$, $O_{uW}^{(i3,3i)}$ and $O_{uG}^{(i3,3i)}$ operators
correspond to weak- and color-dipole couplings.  In particular,
$O_{uG}^{(i3,3i)}$ could induce the production $pp\to th$, and it was not
included in \cite{Wang:2012gp}.  The last operator gives rise to
flavor-changing Yukawa couplings.  This operator is actually implemented as
$O_{u\varphi}^{(i3)}=(\varphi^\dagger\varphi-v^2/2)(\bar{q}_it)\tilde\varphi$
to avoid any need for a field redefinition in order to remove the tree-level
$q-t$ mixing.  It is interesting to note that all $qtB$ interactions receive
contributions from operators that involve the Higgs field, therefore they are
also relevant for constraining new physics in the Higgs sector.  Finally, we
stress that four-fermion operators should  also be taken into account for a
complete phenomenological study of FCN interactions, see
Ref.~\cite{inprogress}.  Their implementation in the current framework is
possible and is left for future work.

In calculations at NLO in QCD, a  renormalization scheme needs to be specified, in
particular for the dimension-six operators.  We adopt the
$\overline{\textrm{MS}}$ scheme in general, except for masses and wave functions
that are renormalized on shell. Specifically, this requires the
introduction of off-diagonal wave function counterterms to cancel the $u-t$ or
$c-t$ two-point functions generated by $O_{uG}^{(i3,3i)}$. We work in the
five-flavor scheme where the $b$-quark mass is neglected, and we subtract the
massless modes according to the $\overline{\textrm{MS}}$ scheme and the top at
zero momentum for the strong coupling constant renormalization
\cite{Collins:1978wz}. At order $\alpha_S$ these operators will not mix with
the SM terms, but mix among themselves.  The running of these coefficients is
given by the renormalization group equations
\begin{equation}
  \frac{\mathrm{d}C_i(\mu)}{\mathrm{dln}\mu}=\gamma_{ij}C_j(\mu)\ ,
\end{equation}
where $\gamma_{ij}$ for $C_{uG}^{(13)}$, $C_{uW}^{(13)}$, $C_{uB}^{(13)}$ and
$C_{u\varphi}^{(13)}$ can be written as a matrix
\cite{Zhang:2014rja,Alonso:2013hga}:
\begin{equation}
  \gamma=\frac{\alpha_S}{\pi}\left(
  \begin{array}{ccccccc}
    \frac{1}{3}	&	0	&	0	&	0\\
    \frac{2}{3}	&\frac{2}{3}	&	0	&	0\\
    \frac{10}{9}	&	0	&\frac{2}{3}	&	0\\
    	4y_t^2	&	0	&	0	&	-2\\
  \end{array}
  \right)
  \ ,
  \label{eq:AD3}
\end{equation}
where $y_t$ is the top-quark Yukawa coupling.  The same $\gamma_{ij}$ matrix
applies for the operators with either $(i3)$ or $(3i)$ superscript.  The operators
$O_{\varphi q}^{(3,i+3)}$, $O_{\varphi q}^{(1,i+3)}$ and $O_{\varphi
u}^{(i+3)}$ do not have any anomalous dimension due to current conservation and
do not mix with other operators.

\section{Implementation and checks}
The operators are implemented in the UFO format \cite{Degrande:2011ua}, using
the {\sc FeynRules} package \cite{Alloul:2013bka}.  The evaluation of the loop
corrections in {\sc MadGraph5\_aMC@NLO} requires two additional elements, the UV
counterterms and the rational R2 terms which are required by the OPP technique
\cite{Ossola:2006us}.  These are computed fully automatically by the NLOCT
\cite{Degrande:2014vpa} package, which has been extended to handle EFT's
\ie,~to compute the R2 and UV divergent parts of amplitudes with
integrals of arbitrary high ranks. Currently, such calculations are limited to
operators with up to two fermion fields. The determination of the UV divergent
part of the counterterms is obtained by simply changing the sign of the UV
divergent part of the corresponding  amplitude.  This avoids the translations
of the counterterms vertices in the operator renormalization constants and the
associated  basis reduction. However, it is only valid when the dimension-six
operators are renormalized in the $\overline{\textrm{MS}}$ scheme.

We have extensively checked our implementation by evaluating the virtual
contributions of $ug\to t$, $u\gamma\to t$, $uZ\to t$, $uh\to t$ and $ug\to th$
(with $uth$ coupling only) and comparing them with corresponding known
analytical expressions numerically.  In each case the results agree.  In
addition we have checked the gauge invariance of all virtual contributions, as
well as the pole cancellation when combining virtual and real contributions.
When possible, we have also made comparisons with the results for total cross
sections for $pp\to t\gamma, tZ, th$ at the fixed order of
Refs.~\cite{Zhang:2011gh,Li:2011ek,Wang:2012gp}, finding consistent results. 

\section{Calculation} 
\begin{figure}[t]
  \begin{minipage}{.49\linewidth}
    \begin{center}
      \includegraphics[width=.9\linewidth]{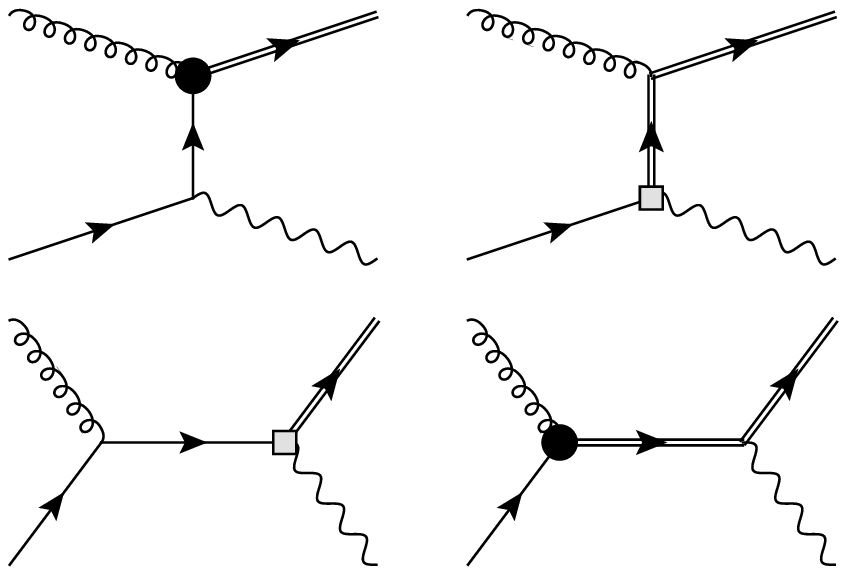}
    \end{center}
  \end{minipage}
  \begin{minipage}{.49\linewidth}
    \begin{center}
      \includegraphics[width=.9\linewidth]{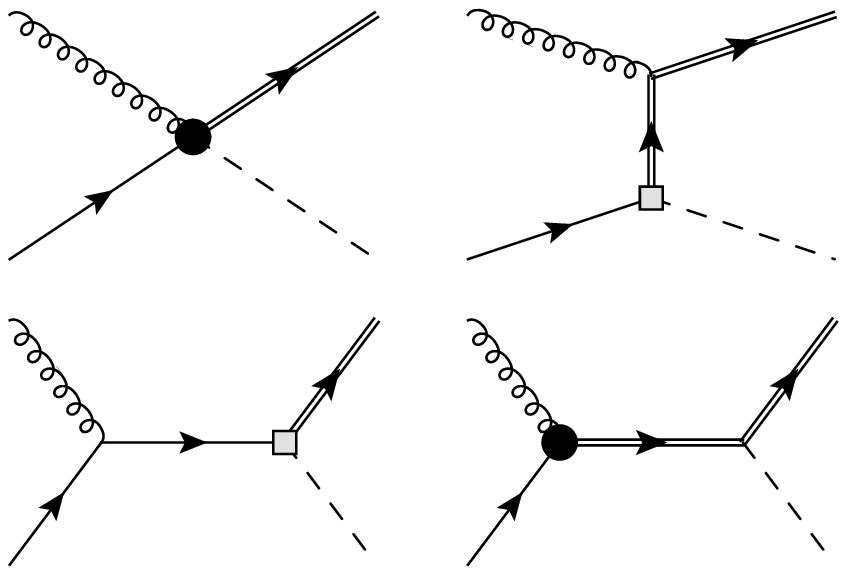}
    \end{center}
  \end{minipage}
  \caption{Tree-level diagrams for $pp\to tV$ and $pp\to th$.
    The black dots represent contributions from color dipole operators
    $O_{uG}^{(i3,3i)}$, while the shaded squares represent other operators.}
  \label{fig:lodiagram}
\end{figure}

As an application of our general framework to the
phenomenology of the top quark FCN at the LHC, we consider three processes,
$pp\to t\gamma$, $pp\to tZ$ and $pp\to th$.  The LO diagrams are shown in
Fig.~\ref{fig:lodiagram}.  Each process receives contributions from two
different interactions,  one from $utg$ coupling and the other from $utB$
coupling.  At NLO in QCD the $utg$ operator will mix with other operators, and
as a result a NLO calculation needs to be carried out with the full set of
operators.

Our numerical results are obtained by  employing the following input parameters
\begin{align}
  &m_Z=91.1876\ \mathrm{GeV},\quad
  \alpha=1/127.9,\quad
  \nonumber\\
  &G_F=1.166370\times 10^{-5}\ \mathrm{GeV}^{-2},
  \nonumber\\
  &
  m_t=172.5\ \mathrm{GeV},\quad
  m_h=125\ \mathrm{GeV},\quad\Lambda=1\ \mathrm{TeV}.
\end{align}
We use CTEQ6M for NLO and CTEQ6L for LO calculations respectively, with their
respective values of $\alpha_S$~\cite{Pumplin:2002vw}.  The renormalization
scale $\mu_r$ and factorization scale $\mu_f$ are chosen to be $m_t+m_B$ for
the $pp\to tB$ process, and are allowed to vary independently by a factor of 0.5 to
2.  In $pp\to t\gamma$, we require the photon $p_T>50$ GeV and its
pseudorapidity $|\eta|<2.5$.  For the photon, we employ the isolation criterium
of Ref.~\cite{Frixione:1998jh} with a radius of 0.4.  The events are then
showered with PYTHIA6 \cite{Sjostrand:2006za} or HERWIG6 \cite{Corcella:2000bw}.
Finally, we have checked that the doubly resonant diagrams with the antitop
decaying through FCN interactions have a small impact, yet they have been
removed from the real contributions, see Ref.~\cite{Frixione:2008yi}. 

Currently the best limits on top FCN couplings are from the decay searches of
$t\to qZ$ \cite{Chatrchyan:2013nwa}, $t\to qh$ \cite{CMS:2014qxa,Aad:2014dya},
and the production searches of $qg\to t$ \cite{TheATLAScollaboration:2013vha}
and $qg\to t\gamma$ \cite{CMS:2014hwa}.  To make a viable choice for
the operator coefficients in our calculation, we exploit the results of
Ref.~\cite{Durieux:2014xla} that are based on a global fit on the full set of
current limits
\begin{equation}
  \begin{array}{|c|c||c|c||c|}
    \hline
    \mathrm{Coefficient} & \mathrm{Limit}&
    \mathrm{Coefficient} & \mathrm{Limit}
    & {}^{\mbox{Relevant}}_{\mbox{production}}
    \\\hline
    C_{\varphi q}^{(j,i+3)} & 1.05
    &
    C_{\varphi u}^{(i+3)} & 1.05
    & tZ
    \\\hline
    C_{uG}^{(13,31)}&0.041
    &
    C_{uG}^{(23,32)}&0.093
    &t\gamma,tZ,th
    \\\hline
    C_{uW}^{(13,31)}&0.92
    &
    C_{uW}^{(23,32)}&1.1
    &t\gamma,tZ
    \\\hline
    C_{uB}^{(13,31)}&1.0
    &
    C_{uB}^{(23,32)}&1.9
    &t\gamma,tZ
    \\\hline
    C_{u\varphi}^{(13,31)}&3.5
    &
    C_{u\varphi}^{(23,32)}&3.5
    &th
    \\\hline
  \end{array}
  \nonumber
\end{equation}
where $i=1,2$, $j=1,3$, and the limits apply to the moduli of the
coefficients, assuming $\Lambda=1$ TeV.  Each limit is obtained by
marginalizing over all the other operator coefficients. In this work, we choose
real and positive values for the coefficients that do not exceed these bounds.
The total cross sections at the LHC at $\sqrt{s}=13$ TeV corresponding to each
operator are displayed in Tables~\ref{tab:ppta}, \ref{tab:pptz} and
\ref{tab:ppth}.  The scale uncertainties are also displayed.  As expected
the $K$ factors are generally sizeable and the scale uncertainties are
significantly reduced at NLO. This is the case for all operators except for
$O_{uG}^{(i3,3i)}$ in $t\gamma$ production. This process has an unusually large
$K$ factor when the flavor-changing coupling is coming from $O_{uG}^{(i3,3i)}$.
As shown in Table~\ref{tab:ppta}, vetoing any extra jet with $p_T>50$ GeV
reduces the $K$ factor from 2.3 (3.3) to 1.6 (2.3) for $utg$ ($ctg$) coupling
as well as the uncertainties for this production mechanism.  Note also that a
jet veto can help to improve the signal over the SM background ratio, for all
three processes.

\begin{table}
  \centering
  \begin{tabular}{|c|c|c|c|c|}
    \hline
    &\multicolumn{2}{|c|}{LO}&\multicolumn{2}{|c|}{NLO}
    \\\hline
    Coefficient & $\sigma$[fb] & Scale uncertainty & $\sigma$[fb] & Scale uncertainty
    \\\hline
    $C_{uB}^{(13)}=1.0$ & 546 & +14.4\% -11.8\%& 764 &+6.9\% -6.4\%
    \\\hline
    $C_{uG}^{(13)}=0.04$ & 1.00 & +12.0\% -10.2\%& 2.34 &+15.2\% -11.5\%
    \\\hline
    $C_{uG}^{(13)}$, veto & 0.739 & +11.50\% -9.8\%& 1.19 &+7.7\% -6.5\%
    \\\hline
    $C_{uB}^{(23)}=1.9$ & 152 & +10.6\% -9.6\%& 258 &+6.8\% -6.0\%
    \\\hline
    $C_{uG}^{(23)}=0.09$ & 0.590 & +12.1\% -11.1\%& 1.95 &+16.4\% -12.3\%
    \\\hline
    $C_{uG}^{(23)}$, veto & 0.457 & +12.2\% -11.2\%& 1.04 &+10.3\% -8.9\%
    \\\hline
  \end{tabular}
  \caption{Total cross sections for $pp\to t\gamma$.  Contributions from
    operators with (31), (32) superscripts are not displayed, but
    they are the same as their (13), (23) counterparts.  Contributions from
    $O_{uW}^{(i3),(3i)}$ are equal to those from $O_{uB}^{(i3),(3i)}$.}
  \label{tab:ppta}
\end{table}
\begin{table}
  \centering
  \begin{tabular}{|c|c|c|c|c|}
    \hline
    &\multicolumn{2}{|c|}{LO}&\multicolumn{2}{|c|}{NLO}
    \\\hline
    Coefficient & $\sigma$[fb] & Scale uncertainty & $\sigma$[fb] & Scale uncertainty
    \\\hline
    $C_{\varphi u}^{(1+3)}=1.0$ & 905 & +12.9\% -10.9\%& 1163 &+6.2\% -5.6\%
    \\\hline
    $C_{uW}^{(13)}=0.9$ & 1737 & +11.5\% -9.8\%& 2270 &+6.6\% -6.2\%
    \\\hline
    $C_{uG}^{(13)}=0.04$ & 30.1 & +17.5\% -13.8\%& 36.0 &+3.8\% -5.2\%
    \\\hline
    $C_{uG}^{(31)}=0.04$ & 29.4 & +17.7\% -13.9\%& 34.9 &+3.4\% -5.1\%
    \\\hline
    $C_{\varphi u}^{(2+3)}=1.0$ & 73.2 & +10.4\% -9.3\%& 107 &+6.5\% -5.9\%
    \\\hline
    $C_{uW}^{(23)}=1.1$ & 172 & +7.5\% -7.2\%& 255 &+6.1\% -5.2\%
    \\\hline
    $C_{uG}^{(23)}=0.09$ & 6.92 & +11.3\% -9.9\%& 10.6 &+5.8\% -5.4\%
    \\\hline
    $C_{uG}^{(32)}=0.09$ & 6.58 & +11.5\% -10.1\%& 10.0 &+5.7\% -5.3\%
    \\\hline
  \end{tabular}
  \caption{Total cross sections for $pp\to tZ$.  Contributions from operators
    $O_{uW}^{(31),(32)}$ are the same as those from $O_{uW}^{(13),(23)}$.
    Contributions from $O_{uB}^{(i3),(3i)}$ are equal to those from
    $O_{uW}^{(i3),(3i)}$ times $\tan^4\theta_W$.  Contributions from
    $O_{\varphi q}^{(j,i+3)}$ are the same as those from $O_{\varphi
    u}^{(i+3)}$.}
  \label{tab:pptz}
\end{table}
\begin{table}
  \centering
  \begin{tabular}{|c|c|c|c|c|}
    \hline
    &\multicolumn{2}{|c|}{LO}&\multicolumn{2}{|c|}{NLO}
    \\\hline
    Coefficient & $\sigma$[fb] & Scale uncertainty & $\sigma$[fb] & Scale uncertainty
    \\\hline
    $C_{u\varphi}^{(13)}=3.5$ & 2603 & +13.0\% -11.0\%& 3858 &+7.4\% -6.7\%
    \\\hline
    $C_{uG}^{(13)}=0.04$ & 40.1 & +16.5\% -13.2\%& 50.7 &+4.0\% -5.2\%
    \\\hline
    $C_{u\varphi}^{(23)}=3.5$ & 171 & +9.7\% -8.7\%& 310 &+7.3\% -6.3\%
    \\\hline
    $C_{uG}^{(23)}=0.09$ & 9.53 & +11.0\% -9.7\%& 16.6 &+5.5\% -5.1\%
    \\\hline
  \end{tabular}
  \caption{Total cross sections for $pp\to th$.  Contributions from
    operators $O_{u\varphi}^{(3i)}$ and $O_{uG}^{(3i)}$ are equal to those from
    $O_{u\varphi}^{(i3)}$ and $O_{uG}^{(i3)}$, respectively.}
  \label{tab:ppth}
\end{table}

\section{Differential cross sections} 
\begin{figure}[tb]
  \begin{center}
  \includegraphics[width=.9\linewidth]{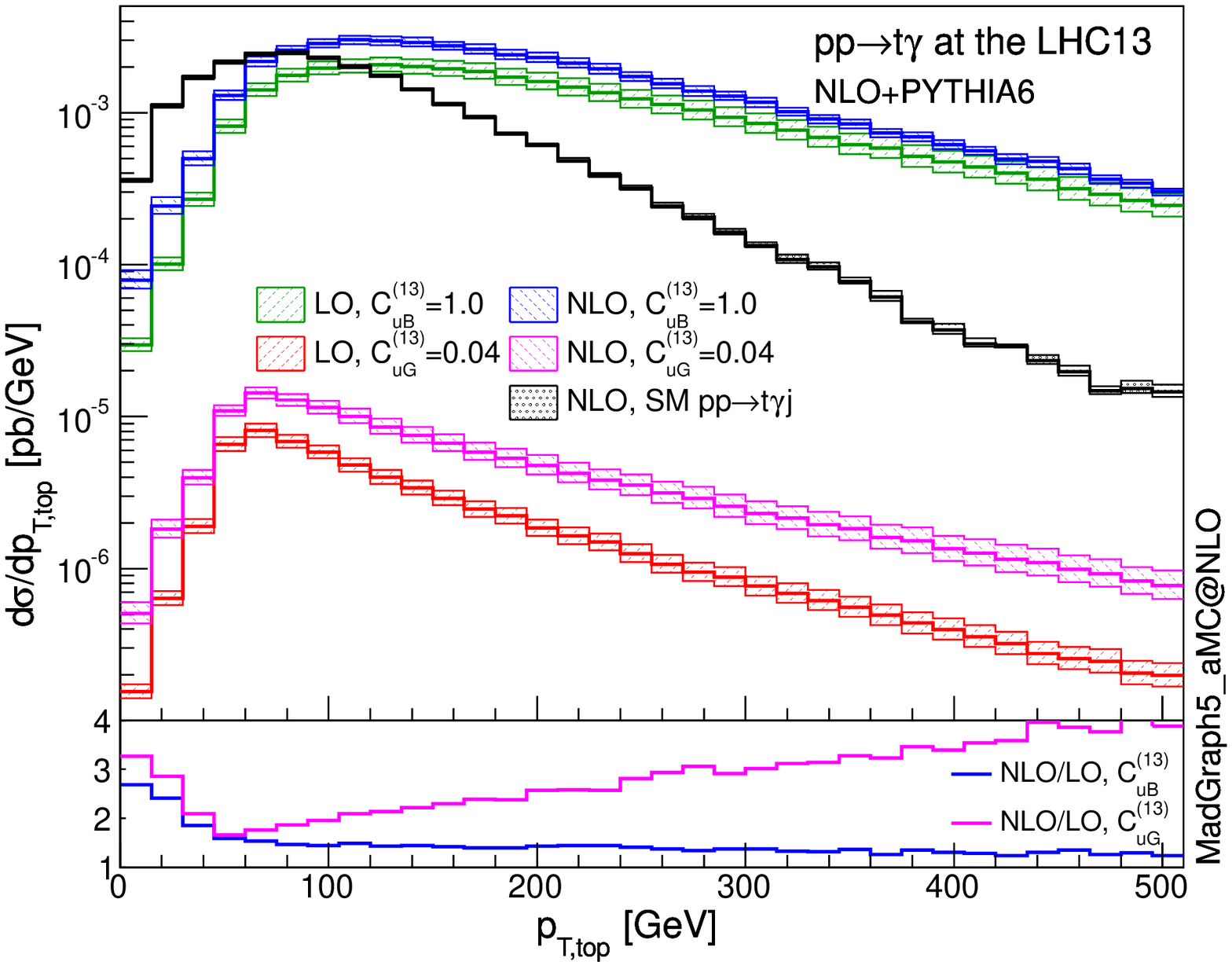}
    \includegraphics[width=.9\linewidth]{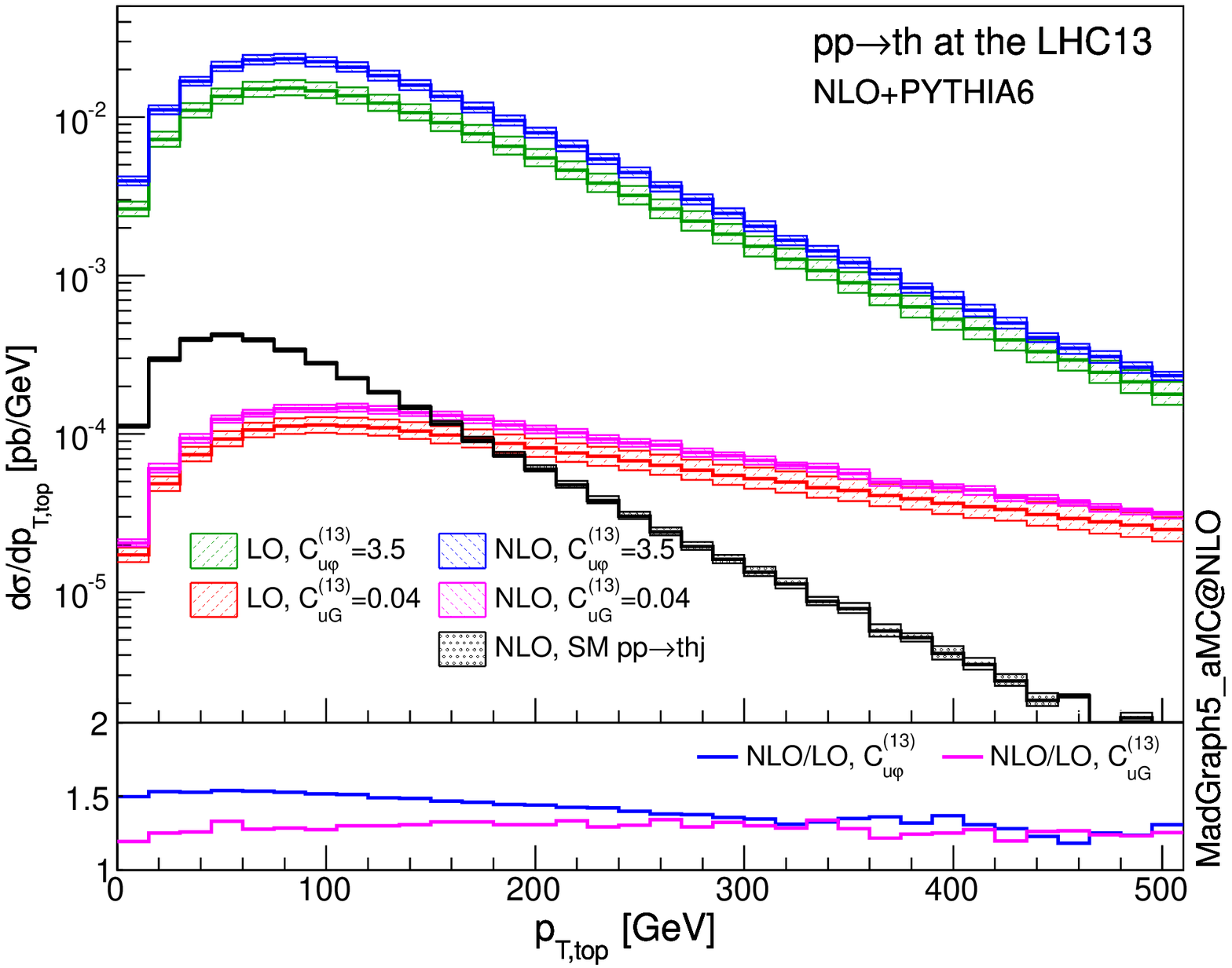}
  \end{center}
  \caption{The $p_T$ distribution of top quark in $pp\to t\gamma$ (top)
    and in $pp\to th$ (bottom).}
  \label{fig:pt}
\end{figure}

The $p_T$ distributions of the top quark
in $pp\to t\gamma$ and $pp\to th$ are shown in Fig.~\ref{fig:pt}.  Both LO and
NLO signals are displayed, together with the SM backgrounds from $pp\to t\gamma
j$ and $thj$, which are generated at NLO with the same parameters.  
In all cases the $O_{uG}^{(13)}$ contributions are very small due to the
stringent limit from $ug\to t$ production. Therefore, the $p p \to t X$
processes appear more as confirmation than as a discovery channel for the
chromomagnetic operator.

\begin{figure}[thb]
  \begin{center}
    \includegraphics[width=0.9\linewidth]{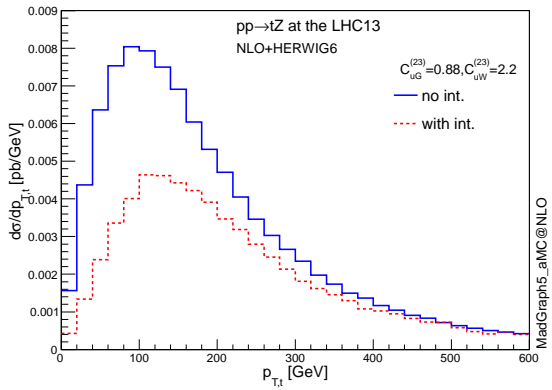}
  \end{center}
  \caption{The interference effect in $pp\to tZ$, as a function of top $p_T$.}
  \label{fig:int}
\end{figure}

To illustrate the importance of keeping all operators possibly contributing to
a given final state, we illustrate in Fig.~\ref{fig:int}  the interference
effect between $O_{uW}^{(23)}$ and $O_{uG}^{(23)}$, in $pp\to tZ$ production.
As a matter of fact, the interference between these two operators is large and
gives rise to a significant change in the rate as well as in the distributions.

\begin{figure}[thb]
  \begin{center}
    \includegraphics[width=0.9\linewidth]{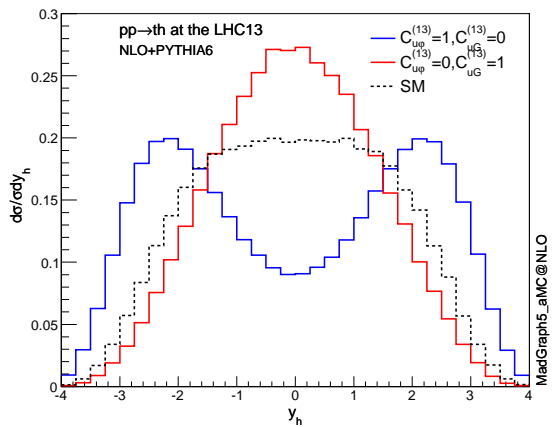}
  \end{center}
  \caption{Normalized rapidity distribution of the Higgs boson in
    $pp\to th$ induced by operators $O_{u\varphi}^{(13)}$ and $O_{uG}^{(13)}$.}
  \label{fig:shape}
\end{figure}

Finally, Fig.~\ref{fig:shape} shows an example where kinematic variables can be
used to distinguish the contributions between different operators.  The Higgs
boson rapidity distribution in $pp\to th$ for $tuh$ coupling induced production
is more forward than that induced by the $tug$ coupling.  The reason is that an
incoming up quark, which is in general more energetic than a gluon, can emit a
forward Higgs boson and turn into an off-shell top quark via a $uth$
vertex, while the same mechanism is not possible for the $utg$ mediated
production.  The same observable may also be used to discriminate between $uth$
and $cth$ couplings, as proposed in Ref.~\cite{Greljo:2014dka}, because $c$ and $g$
have similar PDFs.

\section{Summary} Precision top-quark physics will be one of the priorities
at the next run of the LHC.  The detection of new interactions and in
particular of FCN ones, will be among the most promising searches for new
physics.  A consistent framework to perform such searches is provided by
the dimension-six SM, \ie,~the SM Lagrangian augmented by all operators of
dimension-six compatible with the gauge symmetries of the SM. Bounding
the coefficients of such operators first (and possibly determining
them in case of deviations) requires accurate predictions for
both SM and beyond processes. NLO accurate
predictions in QCD are required in order to correctly extract FCN couplings
from measurements or to set reasonable limits on their sizes.  In this paper we
have presented the first complete NLO computation for the single top
production processes $u(c)+g\to t+B$, $B=\gamma,Z,h$, \ie, including all
dimension-six flavor-changing (two fermion) operators.  In particular, the
chromomagnetic operators with their extra nontrivial effects have been added.
Our computation is based on the {\sc MadGraph5\_aMC@NLO} framework, and thus the
computation is fully automatic and can be applied to other FCN processes.  The
matching of the NLO results to the parton shower is included as well.  The
$K$ factors in all the FCN processes are found to be large, and are in general
not constant over the phase space.  Our work is a first step toward the
automation of NLO computations relevant for searches of new interactions
through the effective field theory framework. 

\section{Acknowledgments}
We would like to thank G.~Durieux, R.~Frederix, V.~Hirschi, O.~Mattelaer and
Y.~Wang for many discussions.  C.~D.~is a Durham International Junior Research
Fellow.  This work has been performed in the framework of the ERC Grant No.~291377
``LHCTheory'' and of the FP7 Marie Curie Initial Training Network MCnetITN
(PITN-GA-2012-315877). The research of J.~W.~has been supported by the Cluster
of Excellence {\it Precision Physics, Fundamental Interactions and Structure of
Matter} (PRISMA-EXC 1098).  C.~Z.~has been supported by the IISN ``Fundamental
interactions'' convention 4.4517.08, and by U.S.~Department of Energy under Grant
No.~DE-AC02-98CH10886.

\bibliography{bib}
\bibliographystyle{apsrev4-1_title}

\end{document}